\documentclass[a4paper]{article}

\usepackage{ISCSLP2024}
\usepackage{float}
\usepackage{algorithm}
\usepackage{algorithmic}
\title{Quantitative Analysis of Audio-Visual Tasks: An Information-Theoretic Perspective}
\name{Chen Chen$^1$, Xiaolou Li$^2$, Zehua Liu$^2$, Lantian Li$^2$, Dong Wang$^1$}

\address{Center for Speech and Language Technologies, BNRist, Tsinghua University, China$^1$
\\
Beijing University of Posts and Telecommunications, China$^2$
\thanks{This work was supported by the National Natural Science Foundation of China (NSFC) under Grants No.62301075/62171250. L. Li and D. Wang are the corresponding authors.}}
\email{\{chenc21,wangdong99\}@mails.tsinghua.edu.cn, \{lixiaolou,liuzehua,lilt\}@bupt.edu.cn}

\begin{document}

\maketitle
\begin{abstract}

In the field of spoken language processing, audio-visual speech processing is receiving increasing research attention. Key components of this research include tasks such as lip reading, audio-visual speech recognition, and visual-to-speech synthesis. Although significant success has been achieved, theoretical analysis is still insufficient for audio-visual tasks. This paper presents a quantitative analysis based on information theory, focusing on information intersection between different modalities. Our results show that this analysis is valuable for understanding the difficulties of audio-visual processing tasks as well as the benefits that could be obtained by modality integration. 
  
\end{abstract}
\vspace{-2mm}
\section{Introduction}
\vspace{-1mm}
Human perception of the world is inherently multimodal, with auditory and visual modalities serving as primary information sources. 
The textual modality also plays a crucial role by recording and conveying semantic information in written form. 
Previous research in spoken language processing has predominantly focused on the perception and production of speech signals, primarily considering text and audio modalities.
In recent years, researchers have increasingly incorporated video as an additional information source. The integration of visual information aims to enhance speech processing tasks, particularly in scenarios where the speech signal is weak or absent. 
This approach has fostered a diverse range of research directions, including lip reading or visual speech recognition~\cite{assael2016lipnet,chung_lip_2017}, visual-to-speech synthesis (VTS)~\cite{mira2022svts}, audio-visual speech recognition (AVSR)~\cite{shi2022avsr}, audio-visual speech enhancement~\cite{kang_impact_2022},
audio-visual speech separation~\cite{ephrat2018looking}, and audio-visual feature representation~\cite{shi2022avhubert}.

The theoretical foundation for incorporating visual information into multimodal tasks lies in the complementary nature of audio and visual information, from both production and perception perspectives. 
From the production perspective, Gunnar Fant's source-filter model~\cite{fant1971acoustic} describes the mechanism of speech generation, where the excitation signal generated by the vibration of the vocal cords is modulated by the vocal tract system to form the audible speech. In this process, the shape of the lips affects the configuration of the oral and nasal cavities, influencing the vocal tract and resultant speech signal.
From the perception perspective, humans extract semantic information mainly from the speech signal but can also utilize visual cues to refine what has been heard. For instance, the McGurk effect~\cite{mcgurk1976hearing} demonstrates that when auditory information is inconsistent with visual information, human perception of semantic content can be biased. 
However, these theories provide only qualitative conclusions and are insufficient for supporting fine-grained quantitative analysis for audio-visual tasks.

In practice, due to the rapid increase in training data and the fast evolution of deep learning techniques, the performance of some audio-visual tasks has greatly improved. For example, the performance of lip reading methods~\cite{chang2024conformervsr}, trained with more than 100k hours of visual-speech data, has reached a word error rate (WER) less than 13\% on the challenging LRS3 dataset~\cite{afouras2018deep}.
Similarly, AVSR methods have significantly improved speech recognition performance in very noisy environments~\cite{shi2022avsr,ma2023auto}.
Unfortunately, most of the achievements depend on end-to-end models trained with large volumes of data, lacking theoretical analysis that can quantify the potential gains with audio-visual integration.
This lack of theoretical foundation leads to blindness in model design and data accumulation. For instance, to what extent does lip movement represent semantic meaning? Shall we use the tandem pipeline that converts lip movement to text and then generates speech, or adopt a direct visual-to-speech production? 

In this work, we conduct a quantitative analysis of the information entropy of each individual modality and the mutual information (MI) among different modalities. This analysis deepens our understanding of the uncertainty and correlation among the auditory, visual, and textual modalities. A key obstacle of the analysis is the difficulty in computing entropy and MI with the audio and video features due to their continuity and high dimensionality.
To tackle this difficulty, we propose a clustering-based approach that discretizes continuous and high-dimensional features into discrete features. Our hypothesis is that if the number of clusters is large, the discrete features can well represent the original continuous features.

The rest of the paper is organized as follows.
Section~\ref{sec:related_work} introduces some studies related to multivariate information methods and the usage of information-theoretical analysis in machine learning. 
Section \ref{sec:method} introduces the design of the information analysis, especially the clustering-based approximation approach. Section \ref{sec:experiment} presents the experiments and results on three datasets: CNVSRC-Multi, GRID and LRS3. Section \ref{sec:conclusion} concludes the whole paper.

\vspace{-2mm}
\section{Related Works}
\label{sec:related_work}

\subsection{Multivariate Information Analysis}

According to Shannon's information theory, the amount of information involved in a variable can be represented by the uncertainty it exhibits through probability distribution. 
The amount of information is quantified by information entropy~\cite{shannon1948mathematical}, defined as follows if the variable is discrete:

\begin{equation}
\label{eq:h}
H(x) = - \sum_x P(X=x) \log [P (X=x)]
\end{equation}

The shared information of two variables, or information intersection, is defined by mutual information (MI) in the following form:

\begin{equation}
  \begin{aligned}
    I(X_1;X_2) & = H(X_1) - H(X_1|X_2)         \\
           & = H(X_2) - H(X_2|X_1)         \\
           & = H(X_1) + H (X_2) - H(X_1;X_2)
  \end{aligned}
  \label{eq:mi}
\end{equation}

\noindent where $H(X_1;X_2)$ denotes the joint entropy of two variables $X_1$ and $X_2$, formulated as follows:

\begin{equation}
H(x) = - \sum_{x_1, x_2} P(X_1=x_1, X_2=x_2) \log [P (X_1=x_1, X_2=x_2)].
\end{equation}

MI is widely used to measure the amount of information shared by two variables.
However, when extending to more than two variables, the shared information is not such intuitive. Srinivasa et al.~\cite{Srinivasa2006ARO} defined multivariate mutual information (MMI) as follows:

\begin{equation}
  \begin{aligned}
    I(X_1;X_2;\dots;X_N) =& I(X_1;X_2;\dots;X_{N-1}) \\
                          &-I(X_1;X_2;\dots;X_{N-1}|X_N)
  \end{aligned}
  \label{eq:mmi_inf}
\end{equation}

\noindent It is evident that this definition is recursive, and in each recursive step, a single variable is taken to intersect with the variables already under consideration.



Note that there are other definitions of MMI. For instance, Qiu et al.~\cite{Qiu2012MMI} defined MMI by grouping the variables into an input group and an output group, and using the group MI to represent the variable MMI.
Nicholas et al.~\cite{Nicholas2014synergy} provided a comprehensive summary of various ways of defining information relationships among multiple variables, including synergy, interaction information (II), co-information (CI), and total correlation (TC). 
His definition of CI is essentially identical to Eq.~(\ref{eq:mmi_inf}).


\subsection{Information-Theoretical Analysis in Machine Learning}

Information theory plays a crucial role in machine learning, particularly in the deep learning era. This theoretical framework has greatly enhanced our understanding of neural networks' behaviour and improved their practical performance.

Firstly, entropy-based loss functions are widely used in training neural networks. For example, cross-entropy and its variants (e.g. binary cross-entropy) are perhaps the most commonly used loss functions in classification tasks~\cite{mannor2005cross}. At the same time, KL and JS divergence play central roles in VAE~\cite{kingma2013auto}, GAN~\cite{goodfellow2020generative}, and other models.

MI is also widely used in machine learning. Since MI directly measures the amount of shared information between two variables, minimizing or maximizing MI can serve as a way to implement information regularization. For example, minimizing MI between different representations enforces disentangled features, benefiting multiple tasks involving information manipulation, such as voice conversion, which requires retaining speech content while changing speaker timbre~\cite{wang2021vqmivc}.
Moreover, the information bottleneck (IB) method, which combines minimizing the input-to-feature MI and maximizing feature-to-output MI, has found a wide range of applications involving information filtering~\cite{liu2024learning}.
MI is also used to define training objectives in self-supervised learning (SSL).
For example, Liu et al.~\cite{liu2024revisiting} defined the training objective of SSL speech representation learning as maximizing the MI between different views or parts of the input. This leads to a unified perspective of various SSL models. Finally, a recent work from the same authors~\cite{liu2024revisiting} showed that MI-based metrics can be used to evaluate the quality of SSL models. Note that existing studies often use bi-variate MI. As we will show shortly, we also use tri-variable MI (i.e., MMI) to compute the shared information among video, audio, and text modalities.
\vspace{-2mm}
\section{Method}
\vspace{-2mm}
\label{sec:method}

\subsection{Entropy-based Quantitative Multimodal Information Model}

We focus on audio-visual processing tasks involving three modalities: audio, video, and text. Our goal is to quantitatively analyze the amount of information (i.e., uncertainty) within each modality and the information intersection between two or three modalities. This leads to an audio-visual information analysis, by which we can answer some interesting questions mentioned in the Introduction section.

First, we use $S$, $V$, and $T$ to represent the variables of Speech, Video, and Text. Then, the information entropy of each modality is denoted by $H(S)$, $H(V)$, and $H(T)$. 
We use Eq.~(\ref{eq:mi}) to compute the shared information of any pairs of variables and 
use Eq.~(\ref{eq:mmi_inf}) to compute the shared information among all three variables. 
Specifically, the tri-variate MMI is computed as follows:

\begin{equation}
  \begin{aligned}
    I(V;T;S) =& H(V) + H(T) + H(S) \\
              &- H(T;V)- H(T;S)- H(V;S) \\
              &+ H(V;T;S)
  \end{aligned}
  \label{eq:mmi_vts}
\end{equation}

Table~\ref{tab:denotes} summarizes the statistical quantities collected for the information analysis. With these quantities, we can analyze the uncertainty of each modality and the relationship between any two or three modalities.

\vspace{-2mm}
\begin{table}[htp]
  \centering
  \caption{Quantities defined and their meanings}
    \vspace{-2mm}
  \begin{tabular}{l|l}
      \toprule
      Denotes    & Meaning                                \\
      \midrule  
      $H(S)$     & Total information of audio              \\
      $H(V)$     & Total information of video              \\
      $H(T)$     & Total information of text               \\
      $I(T;S)$   & Shared information of text and audio    \\
      $I(T;V)$   & Shared information of text and video    \\
      $I(V;S)$   & Shared information of video and audio   \\
      $I(V;T;S)$ & Shared information of three modalities  \\
      \bottomrule
  \end{tabular}
  \label{tab:denotes}
\end{table}

To calculate the entropy of each modality, we first need to extract representations for each modality. Since text is inherently discrete, we directly use phonemes as the units to represent the textual information. For audio and video, we examine both raw features and deep features derived from deep neural networks.

\vspace{-2mm}
\subsection{Clustering-based Entropy Estimation}

The entropy of a discrete random variable $X$ (e.g., the phone unit representing the textual modality) can be calculated simply by Eq.~(\ref{eq:h}). For continuous variables (e.g., the audio and visual features), however, the entropy is infinite in nature. A common practice is to compute the \emph{differential entropy} instead, formulated as follows:
\begin{equation}
  H(X) = -\int_{-\infty}^{\infty} p(x) \log p(x) \, dx
  \label{eq:hx_integration}
\end{equation}

However, we cannot use this formula to compute $H(V)$ and $H(S)$ directly because the exact mathematical form of the distributions is unknown. In particular, they are all high-dimensional, so estimating the probability of high-dimensional data is difficult \emph{per se}. 
Moreover, using continuous features for audio and video modalities poses a fundamental difficulty when computing mutual information that involves the text modality. This is because the text variable is discrete, and the definition of MI between a continuous variable and a discrete variable is ill-posed.

\begin{algorithm}[ht]
\footnotesize
\renewcommand{\algorithmicrequire}{\textbf{Input:}}
\renewcommand{\algorithmicensure}{\textbf{Output:}}
\caption{Clustering-based Entropy Estimation Algorithm}
\label{alg}
\begin{algorithmic}[1]
    \REQUIRE $V \rightarrow [L, d_v]$, $S \rightarrow [L, d_s]$, and $T \rightarrow [L, 1]$;
    \ENSURE $H(V)$, $H(S)$, $H(T)$, $H(T;V)$, $H(T;S)$, $H(V;S)$, and $H(V;T;S)$;
    \STATE Learn K-Means models from $V$ and $S$;
    \STATE Apply K-Means models to $V$ and $S$ to get cluster ids $C_V\rightarrow [L,1]$ and $C_S\rightarrow [L,1]$;
    \STATE Calculate the entropy of discrete distribution $C_V$, $C_S$ and $T$ to get $H(V)$, $H(S)$, and $H(T)$;
    \STATE Calculate the entropy of joint discrete distribution $\{T, C_V\}$, $\{T, C_S\}$, $\{C_V, C_S\}$, and $\{C_V, T, C_S\}$ to get $H(T;V)$, $H(T;S)$, $H(V;S)$, and $H(V;T;S)$;
\end{algorithmic}
\end{algorithm}

To address this problem, we choose to discretize the continuous audio and video features by a simple clustering algorithm as shown in Algorithm~\ref{alg}.
The key idea of the cluster-based method is to cluster continuous high-dimensional representations into discrete units and then assign each feature a label according to which cluster it belongs to. With the discrete labels, the probability distribution can be easily calculated according to the frequency of all the features belonging to each cluster. Based on this discrete probability distribution, the information entropy of the original continuous features can be approximately calculated.
This method was also used by Sun et al.~\cite{sun2023rc} to determine the mutual information between two latent representations of speech signals.

\vspace{-2mm}
\section{Experiments}
\vspace{-2mm}
\label{sec:experiment}

\subsection{Data}

To ensure the reliability of our results, we used multiple datasets and feature extraction methods in our experiments. 
CNVSRC-Multi~\cite{chen2024cnvsrc}, GRID~\cite{cooke2006audio}, and LRS3~\cite{afouras2018deep} were used in the experiments.

CNVSRC-Multi is a newly introduced dataset designed for audio-visual processing research, such as audio-visual speech recognition. It contains synchronized audio and video data along with corresponding text transcriptions. The development set of CNVSRC-Multi includes 20,450 audio-video pairs from 43 different speakers. Each video frame contains only the target speaker’s face, and the corresponding audio contains only the speaker’s voice. All videos are formatted at a frame rate of 25 frames per second, and the audio is sampled at 16 kHz.

GRID and LRS3 are the most commonly used datasets for English visual speech recognition (VSR) and visual-to-speech synthesis (VTS) tasks. The GRID dataset is recorded in a constrained environment, whereas the LRS3 dataset comprises videos from online speeches, such as TED Talks. More information about the two datasets can be found in the original papers~\cite{cooke2006audio,afouras2018deep}.

We adopt the same data pre-processing pipeline described in~\cite{chen2024cnvsrc}, consisting of face recognition and alignment operation, followed by cropping out the lip region.
The provided text transcriptions are in characters. We use the Montreal Forced Aligner (MFA) tool\footnote{https://mfa-models.readthedocs.io/en/latest/index.html} to convert these transcriptions into phonemes. In our experiments, the number of phonemes in Chinese (CNVSRC-Multi) and English (GRID and LRS3) are 217 and 69, respectively.

\begin{table*}[htb!]
  \centering
  \caption{Information entropy and MI with different datasets}
  \begin{tabular}{c|ccccccc}
      \toprule
      Dataset       &    $H(T)$   & $H(S)$  & $H(V)$   & $I(T;V)$  & $I(T;S)$  & $I(V;S)$  & $I(V;T;S)$\\
      \midrule
      CNVSRC-Multi  &    4.0510   & 6.6969  & 3.7831   & 0.6900     & 2.1915    & 1.0726   &  0.2795    \\
      GRID          &    2.7235   & 7.0485  & 6.9768   & 1.4134     & 1.8072    & 3.4692   &  0.9112    \\
      LRS3          &    3.2795   & 5.7971  & 6.3618   & 1.7039     & 1.6663    & 2.7924   &  0.7654    \\
      \bottomrule
  \end{tabular}
  \label{tab:info_cluster_num_datasets}
\end{table*}

\begin{table*}[htb!]
  \centering
  \caption{Information entropy and MI value under different cluster numbers}
  \begin{tabular}{c|ccccccc}
      \toprule
      Cluster Number&    $H(T)$   & $H(S)$  & $H(V)$   & $I(T;V)$  & $I(T;S)$  & $I(V;S)$  & $I(V;T;S)$\\
      \midrule
      100           &    4.0510   & 3.0707  & 0.9934   & 0.2035     & 1.2847    & 0.2581   &  0.1430    \\
      200           &    4.0510   & 2.5717  & 1.4336   & 0.2788     & 0.8512    & 0.3250   &  0.1568    \\
      500           &    4.0510   & 4.1104  & 2.3182   & 0.4176     & 1.3742    & 0.5174   &  0.2123    \\
      1000          &    4.0510   & 5.4220  & 2.9880   & 0.5357     & 1.8031    & 0.7267   &  0.2440    \\
      2000          &    4.0510   & 6.6969  & 3.7831   & 0.6900     & 2.1915    & 1.0726   &  0.2795    \\
      \bottomrule
  \end{tabular}
  \label{tab:info_cluster_num}
\end{table*}

\subsection{Experimental Settings}
\label{sec:exp:set}

For the CNVSRC-Multi dataset, the speech feature extractor used is the HuBERT model~\cite{hsu2021hubert}, released by NPU-ASIP. 
This model was trained on the WenetSpeech dataset, which contains 10,000 hours of Chinese audio signals~\cite{zhang2022wenetspeech}. 
The visual deep feature extractor is the encoder from the VSR baseline system of CNVSRC 2023~\cite{chen2024cnvsrc}, trained on CN-CVS~\cite{chen2023cn} and CNVSRC-Multi~\cite{chen2024cnvsrc}.

For the two English datasets, we used the HuBERT~\cite{hsu2021hubert} and AV-HuBERT~\cite{shi2022avhubert} models released by FAIR to extract deep speech and visual features, respectively. 
The HuBERT model\footnote{https://huggingface.co/facebook/hubert-base-ls960} was trained on the LibriSpeech dataset~\cite{panayotov2015librispeech}, and the AV-HuBERT model\footnote{https://github.com/facebookresearch/av\_hubert} was trained on the VoxCeleb datasets~\cite{Nagrani19vox}. 
For comparison, the raw features are also used to perform the analysis, which involves wave frames for the audio modality and grey pixel images for the visual modality.

To facilitate the calculation of MI and MMI, we defined an identical sampling rate for all three modalities, set at 25 Hz, matching the sampling rate of the video stream. This sampling alignment establishes a straightforward frame-to-frame correspondence between any two modalities, simplifying the calculation of statistics involving more than two modalities.

\vspace{-2mm}
\subsection{Main results}

\begin{figure}[htb!]
  \centering
  \includegraphics[width=1.0\linewidth]{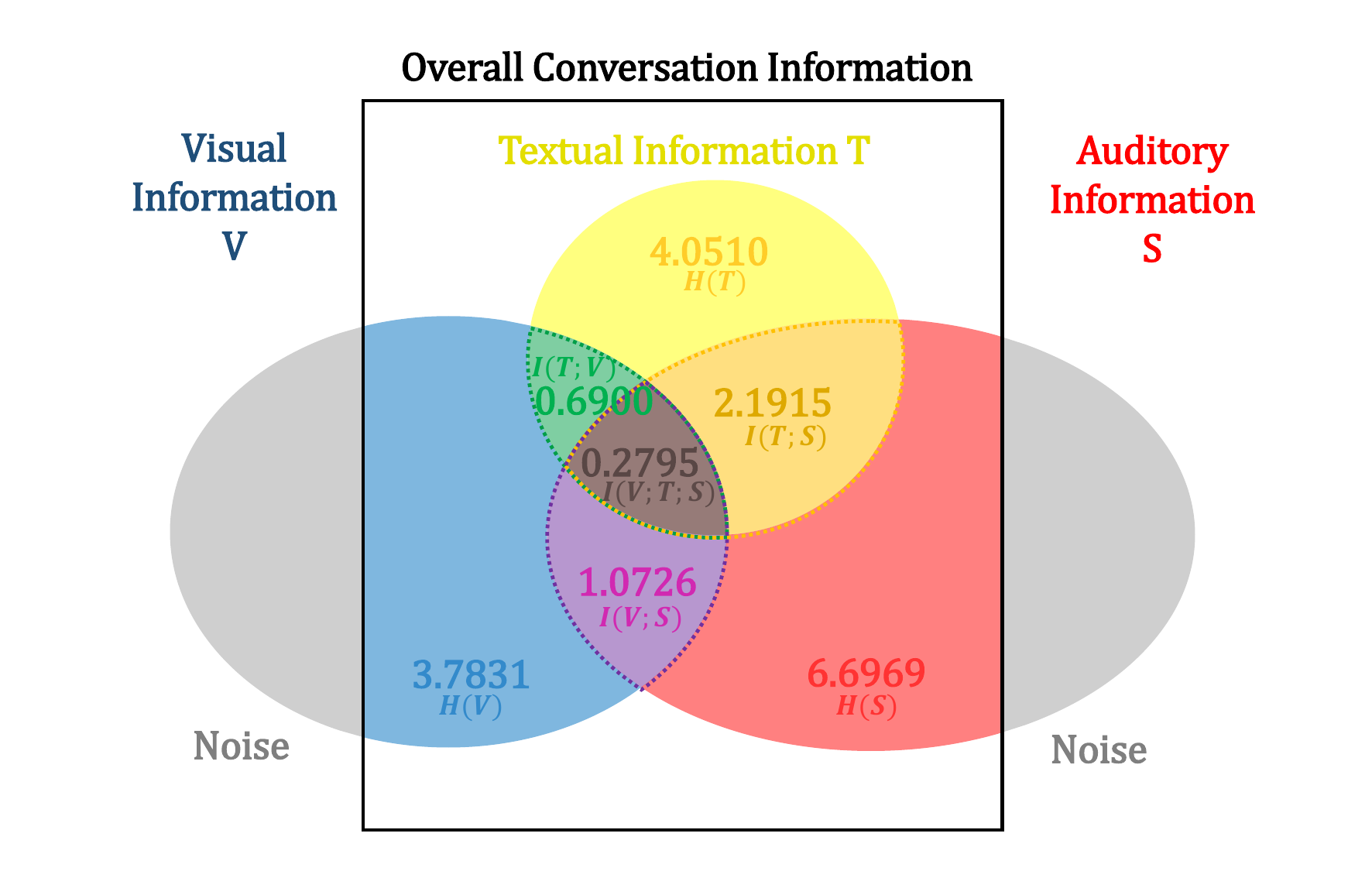}
  \caption{Information diagram computed based on CNVSRC-Multi, using deep features. Note that only the information in the black box is related to the purpose of conversion/speech. The auditory and visual signals partly represent the purpose but also involve some subtle information that is not clearly shown.}
  \label{fig:info_model_nums}
\end{figure}

In the main experiment, we set the number of clusters to 2000 for both the audio and visual modalities and used the deep features mentioned in the previous section. The results on the three datasets (CNVSRC-Multi, GRID, LRS3) are shown in Table~\ref{tab:info_cluster_num_datasets}, where three types of statistics are computed: entropy of each modality, MI between any two modalities, and MMI among the three modalities. For a more vivid representation, Figure~\ref{fig:info_model_nums} presents the information diagram where the statistics are computed on CNVSRC-Multi using deep audio and visual features. 

First, we will focus on the results with the CNVSRC-Multi dataset. It is found that $I(T;V) \ll I(T;S)$ indicates that the speech modality represents more information about the text. Additionally, $I(T;S) \gg I(V;S)$ shows that speech is more related to text than related to video. All these observations imply that the speech modality involves more semantic information than the visual modality. From another perspective, $H(S|T) \gg H(V|T)$ indicates that there is more semantic unrelated information in speech than video, indicating that the speech modality is more complex but more semantically informative than the video modality. 

Moving to the GRID dataset, nearly all the observations are the same as with CNVSRC-Multi, except that $I(V;S)$ is much larger than $I(T;S)$, indicating a close relation between the visual and speech modalities. This seemingly strange observation is consistent with the setting of GRID, where the text is restricted to some simple commands and the recording environment is controlled, so the correlation between visual and speech modalities is significant. 

The results on LRS3 are similar to those on GRID, except that $I(T;V)$ is slightly higher than $I(T;S)$, indicating that the visual modality has similar or even better text predicting capacity than the speech modality. We hypothesize this could be related to the close correspondence between speakers and the content in TED talks, making the text easy to be predicted with both speech and lip movement, especially when the features are derived from a long context.

\subsection{Ablation study}

\subsubsection{Number of clusters}

Given that the number of clusters in the clustering algorithm is closely related to the values of entropy and MI, we show the results with different cluster numbers in Table~\ref{tab:info_cluster_num}.
It can be seen that although the absolute values of the statistics change with the number of clusters, the comparative results remain consistent.
We therefore chose 2000 clusters in the main results. We hope that using a large number of clusters provides an accurate approximation of the original continuous features.

\begin{table}[H]
  \centering
  \caption{Information entropy and MI value with raw and deep features}
  \begin{tabular}{c|cccc}
      \toprule
      Types    & $H(S)$  & $H(V)$   & $I(T;V)$  & $I(T;S)$  \\
      \midrule
      Raw    & 7.2606  & 7.2432   & 0.3615     & 0.5737    \\
      Deep   & 6.6969  & 3.7831   & 0.6900     & 2.1915   \\
      \bottomrule
  \end{tabular}
  \label{tab:info_cluster_num_features}
\end{table}

\subsubsection{Raw and deep features}

We also investigate the statistics with different types of audio and visual features, and the results are shown in Table~\ref{tab:info_cluster_num_features}, where the results are computed on the CNVSRC-Multi dataset, and the raw features are wave frames and grey pixel images for the audio and visual modality, respectively. 
The deep features are extracted as presented in Seciton~\ref{sec:exp:set}. Note that the same number of clusters (2000) is used for the raw and deep features. It can be seen that the entropy values of the raw features are larger than those of deep features, for both the audio and visual modalities. In contrast, the MI values between the raw features and the text variable are much lower than those between the deep features and the text variable. These observations are consistent and reflect the same thing: compared to the raw features, the deep features are less uncertain and more related to the text variable. This is not surprising, as the deep features integrate information of a long context window (especially with Transformers), thus more related to semantic meaning and exhibiting stronger correlation with the text variable.

\section{Conclusions}
\label{sec:conclusion}

We investigated the information load (entropy) and information intersection (mutual information) audio-visual speech processing tasks. 
Through this analysis, we identified different patterns across datasets, but the impact of the features used to conduct the analysis was not significant. For complex datasets like CNVSRC-Multi that involve significant variety in both content and environment, speech conveys more semantic information than video. However, if the content is largely related to speakers, as in the case of LRS3, video can provide reasonable prediction, as demonstrated by the good performance in speaker-dependent lip reading tasks. 
This analysis offers a valuable tool for understanding the behaviour of audio-visual speech processing systems, e.g., the difficulties of speech recognition with individual modalities and the modality complementarity. 

\bibliographystyle{IEEEtran}

\bibliography{mybib}

\end{document}